\documentclass[10pt,twocolumn,journal,compsoc]{IEEEtran}

\usepackage{fancyhdr, epsfig, epsf, amsthm, amsmath, amssymb, amsfonts, subfigure,color}
\usepackage{threeparttable}
\usepackage[noadjust]{cite}
\usepackage{dsfont}
\usepackage{enumerate}
\usepackage{comment}
\usepackage{multirow,booktabs}
\usepackage{bigstrut}
\usepackage[most]{tcolorbox}
\tcbuselibrary{breakable} 
\usepackage{enumitem}
\usepackage{url}

\def\E{\mathsf{E}}

\def\({\left(}
\def\){\right)}

\def\[{\left[}
\def\]{\right]}

\def\papertitle{Distilling On-Device Intelligence at the Network Edge}

\IEEEoverridecommandlockouts

\begin{document}
\title{ \fontsize{18}{25}\selectfont  \papertitle}

\author{Jihong~Park, $^\dagger$Shiqiang~Wang, Anis~Elgabli, $^*$Seungeun~Oh, $^*$Eunjeong~Jeong, $^*$Han~Cha, $^*$Hyesung~Kim, $^*$Seong-Lyun Kim, and Mehdi~Bennis
 \thanks{J.~Park, A.~Elgabli, and M.~Bennis are with the Centre for Wireless Communications, University of Oulu, Oulu 90014, Finland (email: \{jihong.park, anis.elgabli, mehdi.bennis\}@oulu.fi). }
\thanks{$^\dagger$S. Wang is with IBM Thomas J. Watson Research Center, Yorktown Heights, NY, USA (email: wangshiq@us.ibm.com). }
\thanks{$^*$S. Oh, E. Jeong, H. Cha, H. Kim, and S.-L. Kim are with the School of EEE, Yonsei University (email: \{seoh, ejjeong,chan,hskim,slkim\}@ramo.yonsei.ac.kr).}
}

\maketitle \thispagestyle{empty}

\begin{abstract} 
Devices at the edge of wireless networks are the last mile data sources for machine learning (ML). As opposed to traditional ready-made public datasets, these user-generated private datasets reflect the freshest local environments in real time. They are thus indispensable for enabling mission-critical intelligent systems, ranging from fog radio access networks (RANs) to driverless cars and e-Health wearables. This article focuses on how to distill high-quality on-device ML models using fog computing, from such user-generated private data dispersed across wirelessly connected devices. To this end, we introduce communication-efficient and privacy-preserving distributed ML frameworks, termed \emph{fog~ML~(FML)}, wherein on-device ML models are trained by exchanging model parameters, model outputs, and surrogate data. We then present advanced FML frameworks addressing wireless RAN characteristics, limited on-device resources, and imbalanced data distributions. Our study suggests that the full potential of FML can be reached by co-designing communication and distributed ML operations while accounting for heterogeneous hardware specifications, data characteristics, and user requirements.
\end{abstract}
\vspace{-5pt}

\section{Introduction}\label{sec:intro}

Recent advances in communication and computing technologies have created a paradigm shift to fog radio access networks (RANs) in 5G. Fog RAN enables low-latency RAN operations at the network edge, by making use of computing and storing capabilities in wirelessly connected edge devices. Beyond 5G, witnessing recent breakthroughs in machine learning (ML), we envisage that these edge devices will have on-device ML capabilities, thereby locally and promptly carrying out mission-critical RAN operations even when network connectivity is temporarily lost.

There exist two prerequisites for this. One is to utilize user-generated data at the network edge. Edge devices can observe the freshest surrounding environments, so their private datasets are indispensable for mission-critical decisions. The other one is to train on-device ML models collectively across devices over wireless links. Every single device observes a small portion of the global environment, mandating the collaborative training without exchanging privacy-sensitive raw data. These two essential factors call for a novel fog computing based distributed ML framework that is suited for wireless connectivity while preserving local data privacy at the edge, hereafter termed \emph{fog ML (FML)}~\cite{Park:2018aa}.

A notable attempt to enable FML is the federated averaging algorithm~\cite{Brendan17}, also known as \emph{Vanilla federated learning (FL)}. At regular intervals, devices in Vanilla FL exchange model parameters, such as weights and gradients, thereby training their local models collectively without revealing private data. While Vanilla FL preserves privacy, it is still far from reflecting {RAN characteristics} including channel dynamics and network architectures.

\begin{figure}\centering
\includegraphics[width=\columnwidth]{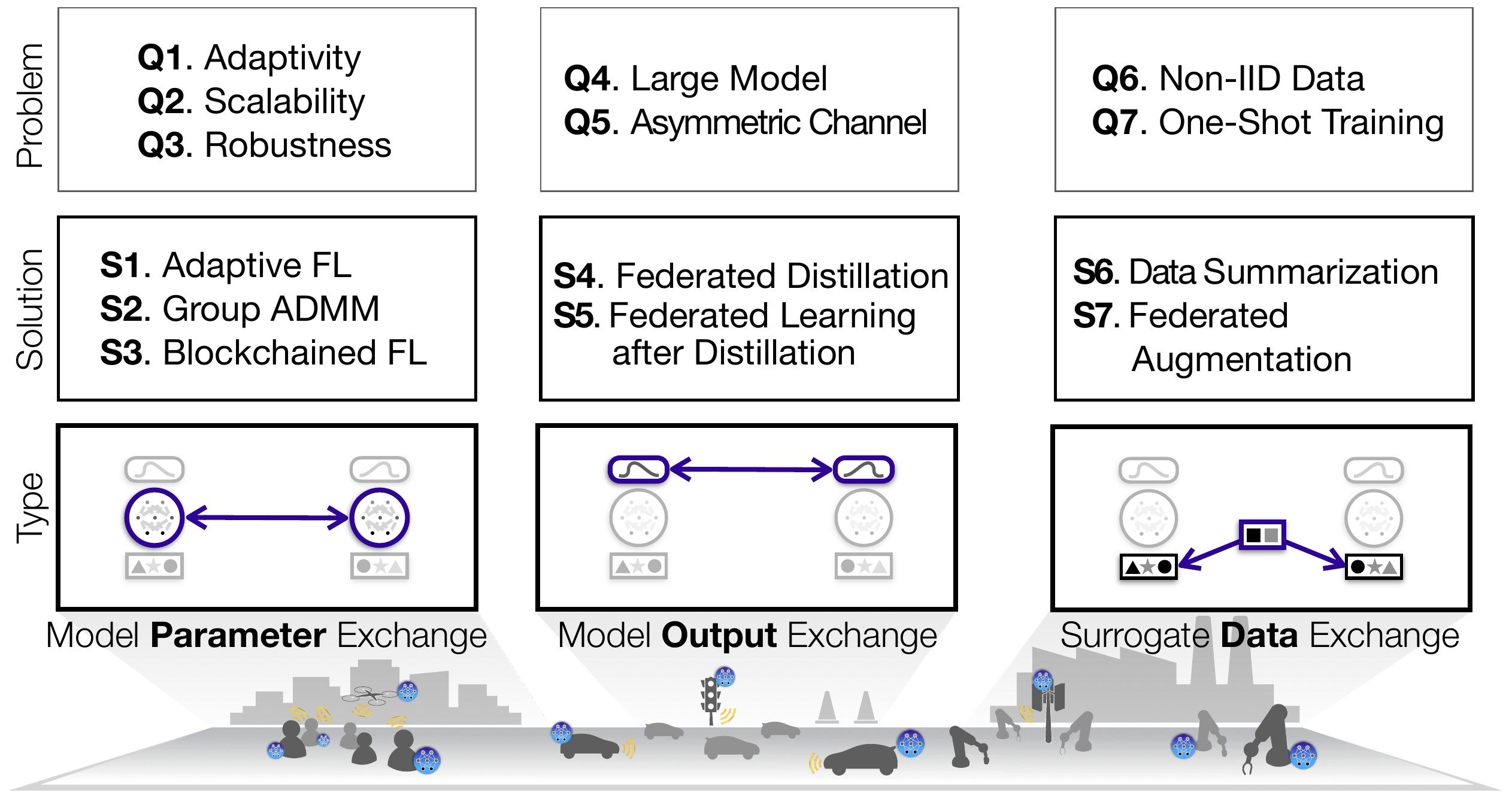} \vskip -7pt
\caption{FML problems \textbf{Q1}-\textbf{Q7} and solutions \textbf{S1}-\textbf{S7}, categorized~into (1) model parameter, (2) output, and (3) surrogate data~exchanges.}\vspace{-5pt}
\end{figure}

\vspace{5pt}\noindent\textbf{Scope and Organization}:
{Towards enabling ML aided fog RAN, this article aims at presenting communication-efficient FML frameworks with high accuracy and data privacy guarantees. To this end, taking Vanilla FL as our baseline, we address the following fundamental questions.}
\begin{tcolorbox}[colback=white, colframe=blue!30!black, boxrule=.5pt, boxsep=3pt,left=1.5pt,right=1.5pt,top=2pt,bottom=2pt, sharp corners]

\noindent \textbf{Q1}. \emph{(Adaptivity)} How often do devices need to communicate for given \emph{channel conditions and training dynamics}? 

\vspace{3pt}
\noindent\textbf{Q2}. \emph{(Scalability)} How to support a large number of federating devices~with \emph{limited transmission power}?

\vspace{3pt}
\noindent\textbf{Q3}. \emph{(Robustness)} How to make FML robust against \emph{malfunctions and adversarial attacks}? \hspace{.8cm}{\small $\Rightarrow$ See \textbf{S1}-\textbf{S3} in Sec.~2}\normalsize

\vspace{-3pt}\noindent\rule{\textwidth}{.5pt}

\vspace{3pt}
\noindent\textbf{Q4}. \emph{(Large Model)} How to deal with large-sized models such as \emph{deep neural networks}?

\vspace{3pt}
\noindent\textbf{Q5}. \emph{(Asymmetric Channel)} How to cope with \emph{asymmetric uplink and downlink channel capacities}? \hfill {\small$\Rightarrow$ \textbf{S4}-\textbf{S5} in Sec. 3}\normalsize

\vspace{-3pt}\noindent\rule{\textwidth}{.5pt}
\vspace{3pt}
\noindent\textbf{Q6}. \emph{(Non-IID Data)} How to rectify datasets that are \emph{not independent and identically distributed (Non-IID)}?

\vspace{3pt}
\noindent\textbf{Q7}. \emph{(One-Shot Training)} How to train FML models in \emph{a single communication round}? {\small\hfill $\Rightarrow$ \textbf{S6}-\textbf{S7} in Sec. 4}\normalsize
\end{tcolorbox}

As shown in Fig.~1, our proposed solutions \textbf{S1}-\textbf{S7} to these problems \textbf{Q1}-\textbf{Q7} boil down to three types of on-device ML model training methods: by exchanging (1) \emph{model parameters}, (2) \emph{model outputs}, and (3) \emph{surrogate data}, which are elaborated respectively in Sec. 2, 3, and 4, followed by conclusion and discussions about extending our proposed FML frameworks in Sec. 5.

\vspace{5pt}\noindent\textbf{Related Works}:
For a comprehensive understanding of FML frameworks, readers are encouraged to check~\cite{Park:2018aa,Yang:2019aa,Vepakomma:18}. Theoretical and algorithmic enablers of on-device ML are overviewed in \cite{Park:2018aa}. Privacy preserving aspects of FML frameworks are studied in \cite{Yang:2019aa,Vepakomma:18}. Note that \cite{Yang:2019aa} provides an FML taxonomy based on data and feature correlations, whereas this article describes FML frameworks from the perspective of \emph{what to exchange} across devices. It is also noted that this work focuses on \emph{communication-efficient data-parallel} architectures wherein devices have an identical model structure but different local data. Model-parallel architectures under which a single model is split across devices are explored in \cite{Vepakomma:18}, ignoring communication efficiency.

\section{Model Parameter Exchange}
 Vanilla FL considers a fixed model averaging interval, and thus cannot cope with instantaneous channel conditions and training dynamics (see \textbf{Q1}). Another drawback is its reliance on a single edge server. The server may not be reachable from faraway devices with low transmission power, due to limited energy or interference mitigation (\textbf{Q2}). Besides, if the server fails due to malfunctions or adversarial attacks, the entire operations will be disrupted, i.e., single point of failure (\textbf{Q3}). To address these problems, this section introduces an advanced FL method adaptive to model training and channel dynamics (\textbf{S1}), as well as server-less FML frameworks with minimum transmission power (\textbf{S2}) and in combination with blockchain (\textbf{S3}).

\subsection{Federated Learning with Adaptive Scheduling}
In Vanilla FL, each device uploads its local model parameters at fixed \emph{communication intervals} to an edge server, i.e., parameter server. By averaging the uploaded parameters, a global model is constructed at the server and downloaded by each device for its local model updates, thereby ensuring robustness against locally unseen data. The communication interval governs not only communication overhead but also accuracy of Vanilla FL. Indeed, reducing communication intervals increases communication overhead, while achieving higher accuracy by better synchronizing local models and avoiding overfitting towards local datasets.

To balance the communication overhead and accuracy, (\textbf{S1}) \emph{Adaptive FL} optimizes its communication interval at the edge server~\cite{Wang:2019aa}, so as to minimize an ML loss function subject to limited communication and computation budgets. It is found that the optimal communication interval depends not only on the communication and computation costs, but also on the data distribution and model characteristics that affect the convexity and smoothness of the loss function. The main building blocks of Adaptive FL are described next.

\begin{figure}\centering
\includegraphics[width=.9\columnwidth]{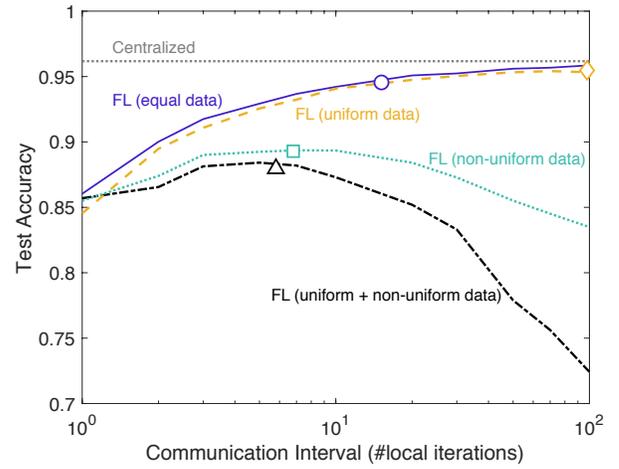}
\caption{
\emph{Adaptive FL} (markers) compared with fixed communication interval baselines (curves) under the MNIST dataset (computation cost : communication cost : total budget = 1:10:1153, see the details in~\cite{Wang:2019aa}).}
\label{fig:adaptiveFL}
\end{figure}

\vspace{5pt}\noindent1) \textbf{Data Distribution Estimation}:
The communication interval should be smaller if the local datasets at different devices are more diverse. To measure this dissimilarity, Adaptive FL evaluates the \emph{gradient divergence} that quantifies the distance between the gradient computed on a local dataset and the average of all such local gradients. A larger gradient divergence indicates more diverse datasets. The gradient divergence can easily be computed in a distributed manner with negligible computation cost, since the model training algorithms rely fundamentally on distributed gradient descent methods.

\vspace{5pt}\noindent2) \textbf{Model Characteristics Estimation}:
Training an ML model is achieved by minimizing a loss function defined on the data samples and model parameters. The model characteristics can therefore be abstracted into the loss function. In view of this, Adaptive FL estimates the loss function's Lipschitzness and smoothness parameters in a distributed fashion, thereby capturing the model properties for the communication interval optimization.

\vspace{5pt}\noindent3) \textbf{Cost Measurement}:
Adaptive FL measures the cost per computing iteration and the cost per communication round, thereby examining whether the total communication and computation budgets are reached. The cost measurements are also used in finding the best communication interval. As an example, for a significant communication cost, a larger communication interval is preferred.

\begin{figure*}\centering
\subfigure[Vanilla FL~\cite{Brendan17}]{\includegraphics[width=.27\textwidth]{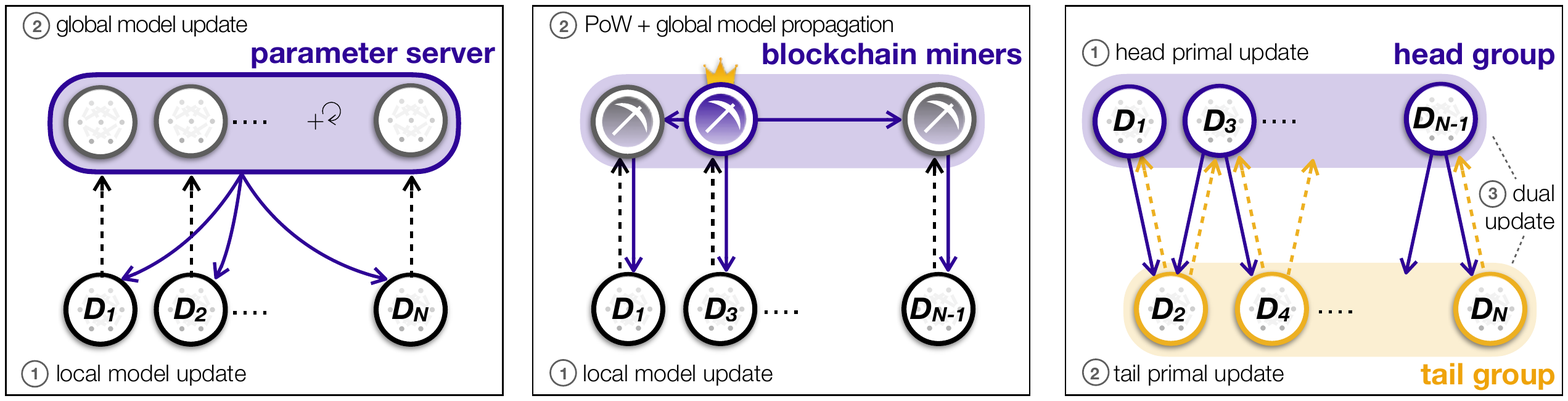}}\hspace{25pt}
\subfigure[GADMM~\cite{Anis:NeurIPS19}]{\includegraphics[width=.27\textwidth]{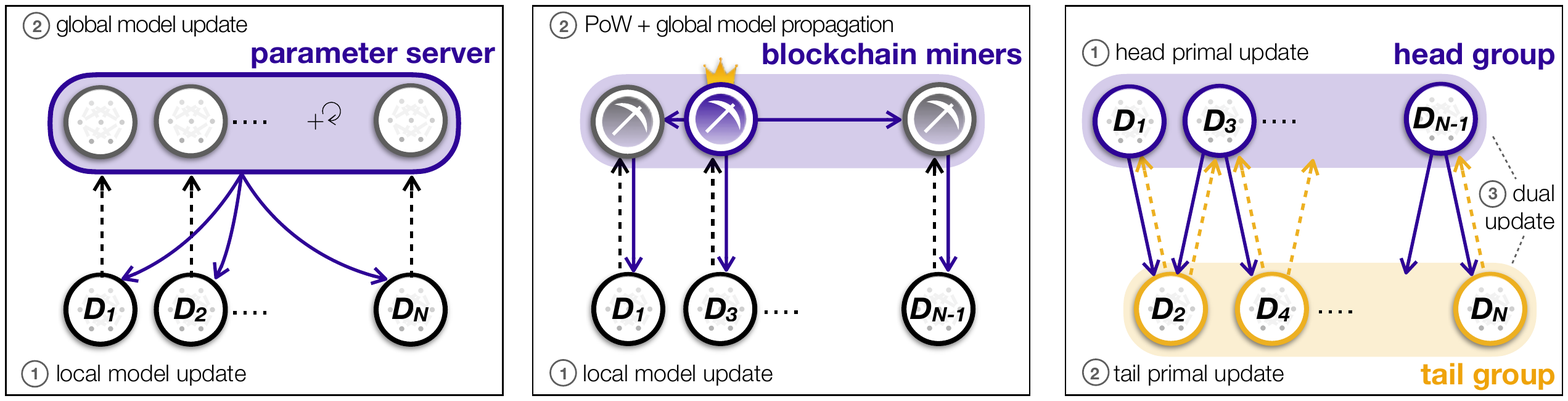}}\hspace{25pt}
\subfigure[BlockFL~\cite{KimCL:19}]{\includegraphics[width=.27\textwidth]{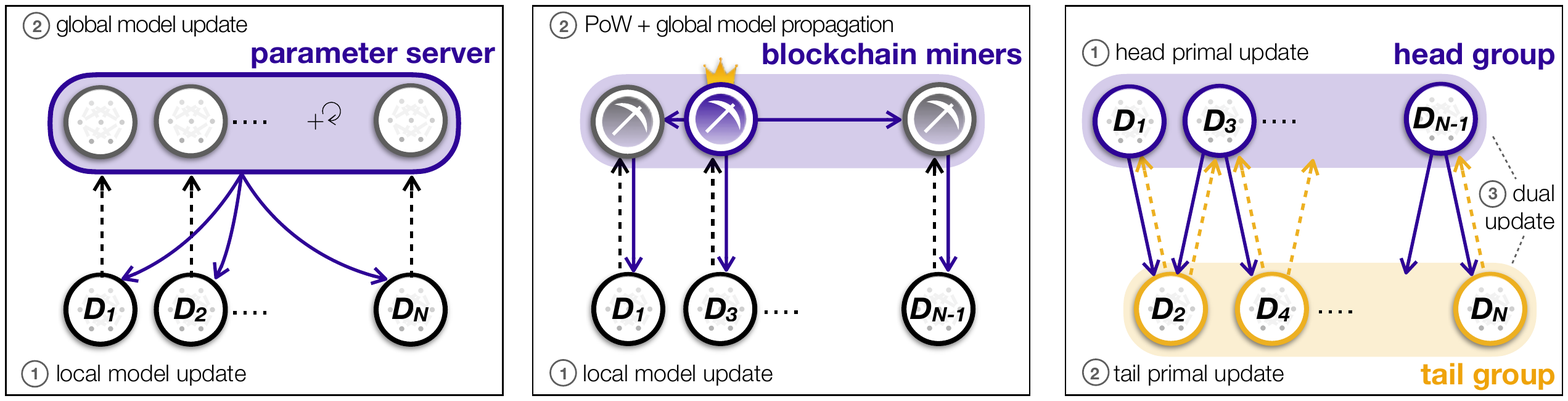}}

\caption{Server-less FML frameworks (b) \emph{GADMM} (Group ADMM) and (c) \emph{BlockFL} (Blockchained FL), compared to the baseline (a) Vanilla FL.}
\label{gadmmFig}
\end{figure*}

Accounting for these three aspects, a closed-form training convergence bound is derived in~\cite{Wang:2019aa}. This bound plays a key role in optimizing the communication interval of Adaptive~FL in a distributed way with low computational complexity. Fig.~\ref{fig:adaptiveFL} compares the test accuracy results under Vanilla FL and Adaptive FL with different data distributions, in which the communication and computation costs are given by real-world measurements in~\cite{Wang:2019aa}. We see that a fixed communication interval as used in Vanilla FL does not work well for all the considered data distributions, because it does not adapt to the data distribution, learning dynamics, and/or cost budget. Furthermore, Fig.~2 shows that the communication interval of Adaptive FL is close to the best achievable interval.

In addition to adjusting the averaging interval in Adaptive FL, communication payload sizes can jointly be optimized under the communication and accuracy trade-off~\cite{Park:2018aa}. To this end, one can adaptively quantize the model parameters to be exchanged by adjusting their arithmetic precision, or exchanging only a fraction of the model parameters, i.e., sparse parameter averaging that is known to achieve the same training convergence rate as periodic averaging~\cite{JiangNIPS2018}.

\subsection{Server-Less FML}

Vanilla FL relies on a single edge server collecting and averaging the model parameters of devices (see Fig. 3a). This server-based architecture is not scalable, as the maximum transmission power of each device is restricted by limited energy and/or inter-device interference. To overcome the limitation, we revisit the optimization problem of Vanilla FL, and present a server-less FML framework as follows.

In brief, Vanilla FL minimizes each device's local loss function, and globally averages their local models, thereby yielding a global model at the server. This can be recast by an average consensus problem that minimizes the loss function averaged across devices, i.e., empirical risk, subject to the constraint making \emph{each local model equal to the global model}. A primal-dual decomposition method can effectively solve this problem in a decentralized way, yet is not scalable since the constraint entails exponentially increasing communication rounds with the number of federating devices.

Alternatively, one can consider a relaxed constraint making \emph{each local model equal only to its two neighboring devices' local models}. (\textbf{S2}) \emph{Group ADMM (GADMM)}~\cite{Anis:NeurIPS19} solves such a modified constrained optimization problem, using the Alternating Direct Method of Multiplier (ADMM) algorithm with a novel device grouping method (see Fig.~3b). GADMM is scalable in a way that it requires each device to communicate only with two neighboring devices, regardless of the total number of federating devices. The operations of GADMM are detailed next.

\vspace{5pt}\noindent1) \textbf{Head-Tail Grouping}:
Devices are divided into two groups \emph{head} and \emph{tail}, such that each head device is connected to other devices through its two neighboring tail devices. The devices in the same group update their model parameters in parallel, while the devices in different groups update the model parameters in an alternating way.

\vspace{5pt}\noindent2) \textbf{Primal Updates}:
At first, each head device locally updates its model parameters, i.e., primal variables. This is done by minimizing the head device's augmented Lagrange function that depends not on other head devices but on its neighboring tail devices' previous primal variables. Then, the updated primal variables are broadcasted to the neighboring tail devices. Next, each tail device updates its primal variables in parallel, based on its neighboring head devices' updated primal variables. Likewise, the updated primal variables are broadcasted to the neighboring head devices. Consequently, at any communication round, only half of the devices broadcast their primal variables.

\vspace{5pt}\noindent3) \textbf{Dual Updates}:
Finally, each device locally updates the dual variables, based on its current primal variables and the primal variables broadcasted from the neighboring devices. Consequently, GADMM achieves the training convergence rate $o(1/k)$ with $k$ iterations, which is faster than $o(1/\sqrt{k})$ of the distributed gradient descent algorithm that commonly forms the basis of Vanilla FL and its variants.

Next, server-less architectures are operated by multiple model parameter aggregating devices. This is useful for avoiding the single point of failure problem, yet makes it challenging to reach a legitimate consensus on a global model. (\textbf{S3}) \emph{Blockchained FL (BlockFL)} resolves this pressing issue by integrating blockchain into FL~\cite{KimCL:19}. In BlockFL (see Fig.~3c), randomly selected devices, i.e., miners, verify and collect the model parameters of the other devices. Then, by solving a random puzzle, i.e., proof-of-work (PoW), the global model is determined by the first winning miner's aggregated model parameters. Furthermore, by leveraging blockchain, BlockFL provides rewards to devices proportionally to their local training data amount, thereby promoting more federation of devices having larger training~data. In general, server-less FML relies solely on low on-device energy and memory, and trading off its accuracy and robustness against communication efficiency subject to each network topology calls for further investigation.

\section{Model Output Exchange}
Model parameter exchange as done in Vanilla FL is ill-suited for large-sized ML models consisting of millions of parameters that consume dozens of gigabytes (see \textbf{Q4}). The problem is aggravated in the uplink whose channel capacity is much lower than downlink capacity (\textbf{Q5}), due to the limited transmission power of devices. Alternatively, this section introduces a model output exchanging FML framework (\textbf{S4}), referred to as \emph{federated distillation (FD)} whose communication payload size is independent of the model size~\cite{Jeong:18}. In combination with FL in the downlink, exploiting FD in the uplink is effective in coping with asymmetric uplink-downlink channels (\textbf{S5}).

\begin{figure}\centering
\includegraphics[width=.9\columnwidth]{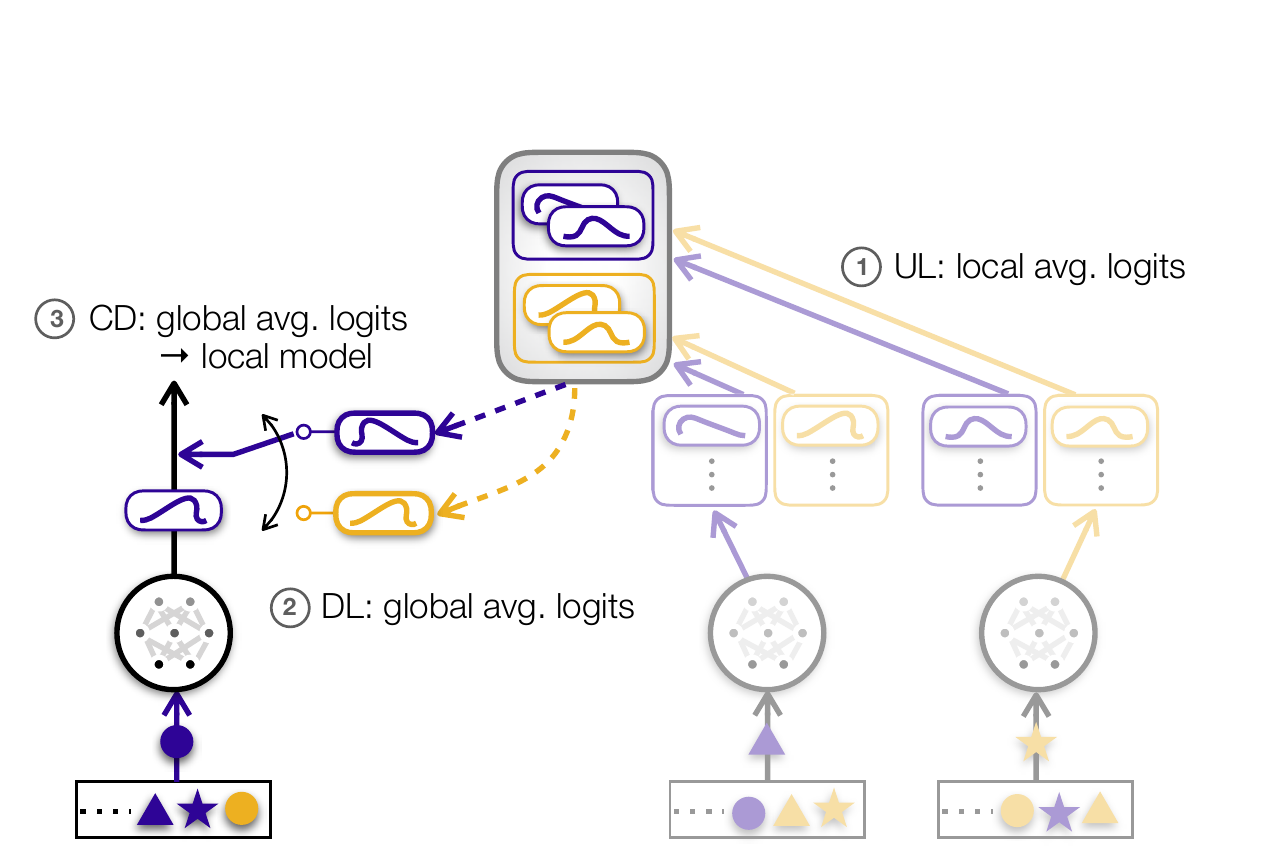}
\caption{\emph{FD} (Federated Distillation): Model output based FML~\cite{Jeong:18}.}
\end{figure}

\subsection{Federated Distillation}
In traditional ML, large-sized models are often handled via pruning and/or reducing the arithmetic precision of model parameters~\cite{Park:2018aa}. These model compression and quantization methods are mostly performed after training, or otherwise achieved at the cost of compromising accuracy. In contrast, we present (\textbf{S4}) FD that reduces the communication cost by exchanging model outputs rather than model parameters, without altering local model structures (see Fig.~4). 

At its core, FD builds on knowledge distillation (KD) that was originally designed for transferring a pre-trained teacher model's knowledge on its prediction capability to an untrained student model~\cite{OnlineKD}. To this end, KD measures the gap between both models' output distributions, i.e., \emph{logits}, of a common data sample, and reflects this gap in updating the student model. Specifically, both teacher and student models in KD first observe a common data sample $x_i$. The student's prediction on $x_i$ is more likely to be wrong, if the student's logit $S_i$ of $x_i$ is less similar to the teacher's logit of $x_i$. In this case, the student increases a penalty in its loss function, i.e., \emph{distillation regularizer}, thereby correcting the student's bias on~$x_i$. FD exploits the said KD operations for FML, while overcoming the following two limitations of KD: (1) \emph{pre-training} of the teacher model and (2) \emph{common sample observations} of teacher and student models. 

FD obviates the problem (1) by adopting co-distillation (CD) \cite{OnlineKD}. The core idea is to make every device become a student, while treating the global average logit $\E[S_i]$ of $x_i$, averaged across devices, as the teacher's logit of $x_i$. However, there still exists the requirement (2). It makes CD exchange the local logits $\{S_i\}$ as many as the number of data samples $\{x_i\}$, incurring huge communication overhead while leaking private information. To resolve this, FD exchanges not local logits of raw samples, but \emph{local average logits} $\{\hat{S}_j\}$ of \emph{proxy samples} $\{\hat{x}_j\}$, as detailed next.

\begin{figure}\centering
\subfigure[FLD]{\includegraphics[width=.9\columnwidth]{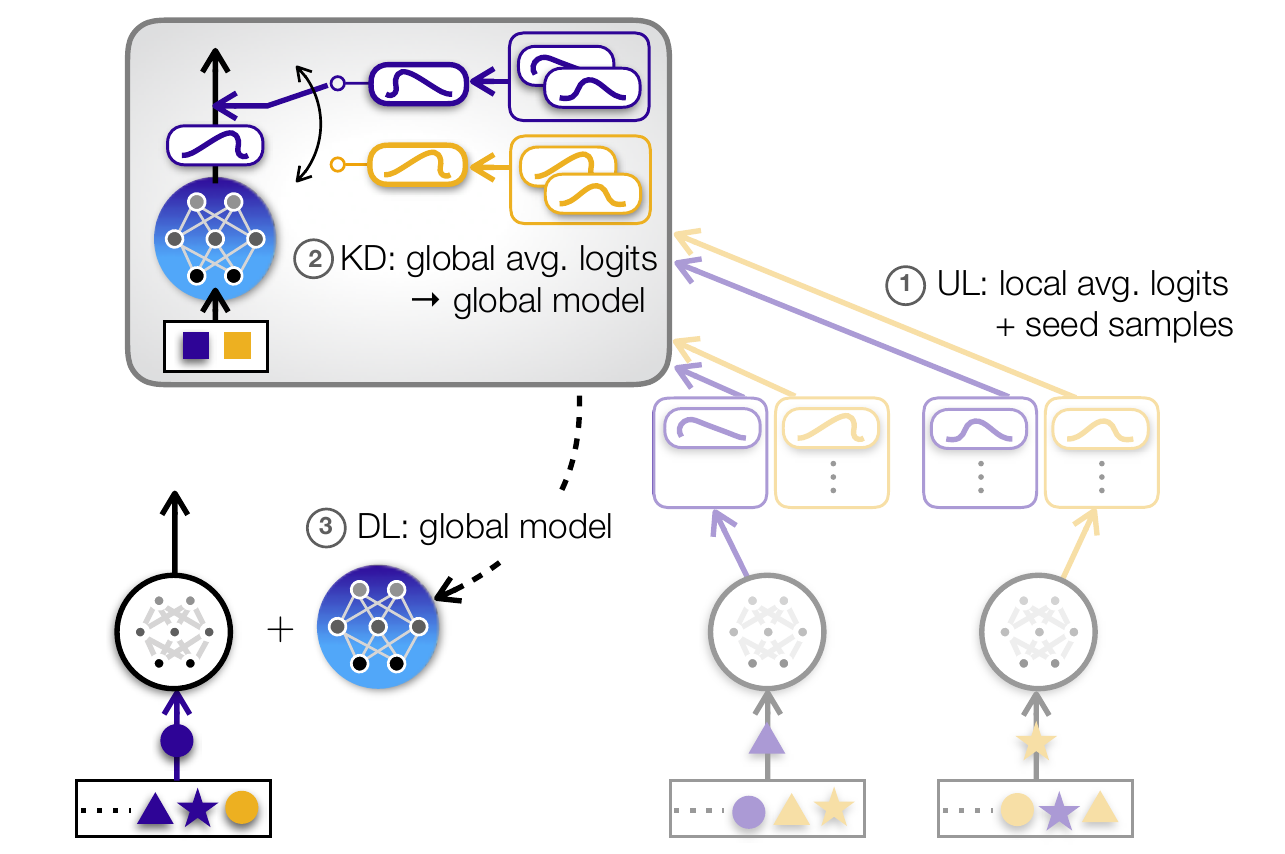}}
\subfigure[Accuracy of FLD, compared to Vanilla FL and FD.]{\includegraphics[width=.95\columnwidth]{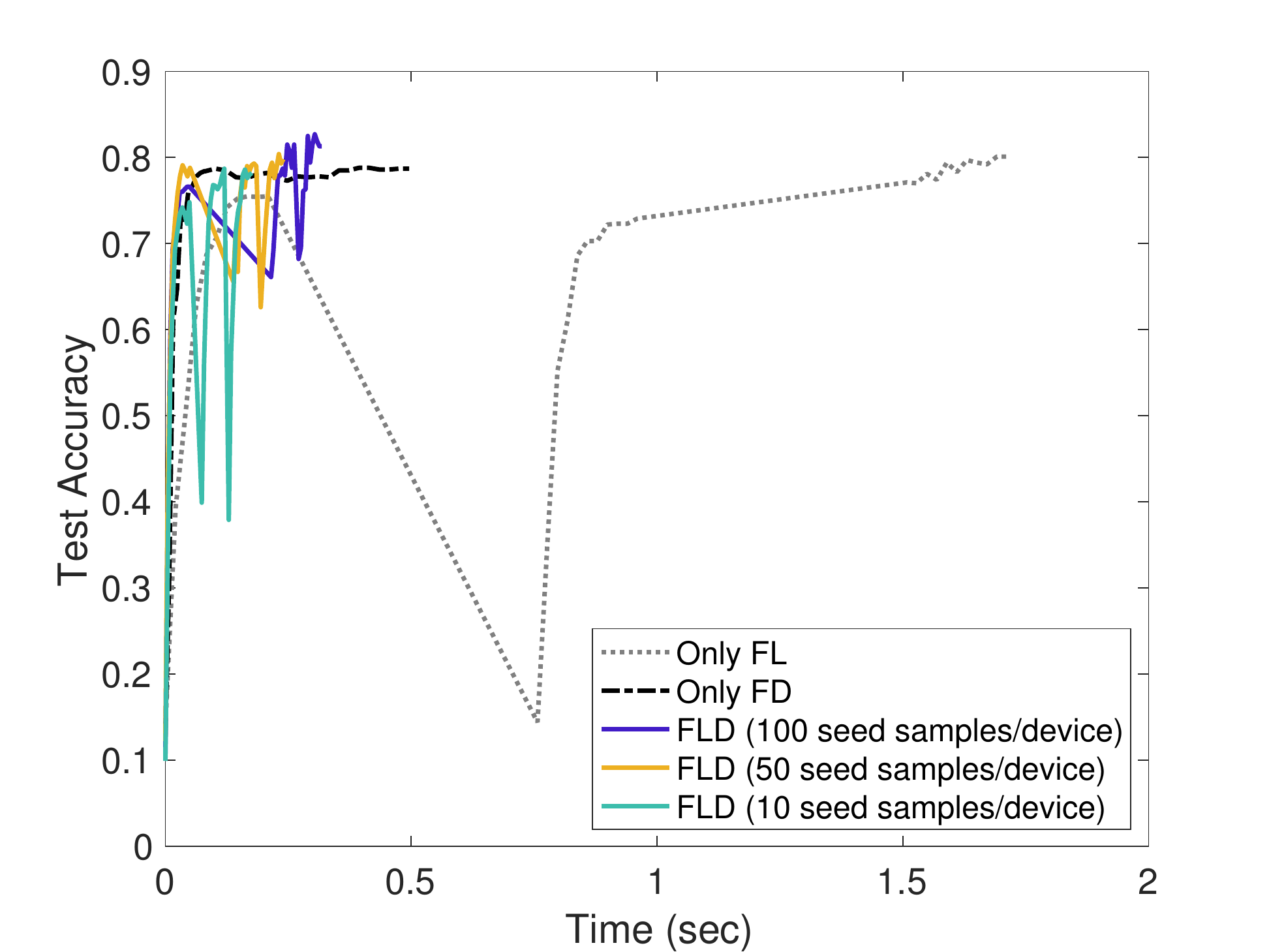}}\vskip -5pt
\caption{\emph{FLD} (Federated Learning after Distillation): Model parameter downloading, after model output uploading from devices and model output-to-parameter converting via KD at the~server.}
\end{figure}

\vspace{5pt}\noindent1) \textbf{FD in Supervised Learning}:
There are four steps in FD operations. Firstly, each device constructs proxy samples fewer than raw samples. In classification tasks, raw samples can be mapped into their ground-truth labels, and one representative raw sample of the ground-truth label is set as a proxy sample~\cite{Jeong:18}. In Fig.~4, for blue and yellow ground-truth labels, two proxy samples $\hat{x}_\text{b}$ and $\hat{x}_\text{y}$ are thereby constructed. Secondly, during local training iterations, each device stores all the logits originated from each ground-truth label separately, and generates local average logits $\hat{S}_{\text{b}}$ and $\hat{S}_{\text{y}}$ at intervals, which are uploaded to the server. Thirdly, by separately averaging the uploaded $\{\hat{S}_\text{b}\}$ and $\{\hat{S}_\text{y}\}$ across devices, the server produces the \emph{global average logits} $\E[\hat{S}_\text{b}]$ and $\E[\hat{S}_\text{y}]$, which are downloaded by devices. Finally, with proxy samples $\hat{x}_\text{b}$ and $\hat{x}_\text{y}$, every device runs CD by comparing its own logits $S_b$ and $S_y$ with the downloaded $\E[\hat{S}_b]$ and $\E[\hat{S}_y]$, respectively, thereby collectively training on-device ML models.

\vspace{5pt}\noindent2) \textbf{FD in Reinforcement Learning}:
In reinforcement learning, quantized states can be proxy states~\cite{Han:FML19}. In the CartPole game, for instance, raw states of angles in $[-180^\circ,180^\circ]$ can be mapped into two quantized angles $\{-45^\circ, 45^\circ\}$, thereby constructing proxy states $\hat{x}_{-45^\circ}$ and $\hat{x}_{45^\circ}$. As done in the supervised learning example 1), each device exchanges local average logits associated with proxy states, referred to as \emph{proxy experience memory}, thereby collectively train on-device ML models. This reduces communication overhead while preserving more privacy, compared to classical distributed reinforcement learning that exchanges raw experience memory containing all logits associated with raw states. A more general approach to constructing proxy samples will be elaborated in Sec.~4.1.

In these ways, FD yields comparable accuracy with Vanilla~FL, while reducing the communication overhead by tenfold, in our experiments with reinforcement learning for completing the CartPole game~\cite{Han:FML19} and supervised learning for classifying the MNIST dataset~\cite{Jeong:18}. The effectiveness of FD may however be limited when local datasets are non-IID across devices. For example, if one device in Fig.~4 only has blue samples while the other device's samples are all yellow, then they cannot transfer any knowledge to each other. A more fine-grained split of proxy states partly resolves this problem, yet may incur more privacy leakage. Therefore, it is essential in FD to rectify non-IID datasets, which will be discussed in Sec.~4.

\subsection{Federated Learning after Distillation}

By marrying FD and FL, we propose (\textbf{S5}) \emph{federated learning after distillation (FLD)}. In FLD (see Fig.~5a), the uplink and downlink operations follow FD and Vanilla FL, respectively. This necessitates a conversion from the uploaded logits to the model parameters to be downloaded. The said conversion is enabled by performing KD at the server lying in between the uplink and downlink. There are two design motivations of FLD as follows.

\vspace{5pt}\noindent1) \textbf{Uplink-Downlink Capacity Asymmetry}:
Uplink capacity is often limited by the low transmission power budget of each device~\cite{JHParkTWC:15}. This may cause a bottleneck in Vanilla FL particularly with large-sized models. In the uplink, it is therefore suitable to exchange logits as done in FD. By contrast, in the downlink with much higher capacity, exchanging model parameters as done in Vanilla FL is preferable. This choice is based on our observations in the MNIST classification experiments~\cite{Jeong:18}, in which FL yields slightly higher accuracy than FD, whereas FD consumes 10x smaller bandwidth.

\vspace{5pt}\noindent2) \textbf{Data Distribution Reflecting Aggregation}: 
FD constructs its global average logits for the downlink, by averaging the local average logits across devices. A simple average cannot account for the data distribution over devices. A weighted average can partly ameliorate this issue, yet the optimal weight selection brings about another challenging problem. By contrast, FLD forms its global model parameters by converting the global average logits to a global model, via KD with few seed samples uploaded from devices. These seed samples are not only required for running KD, but also useful for taking into account the data distribution over devices. 
\vspace{5pt}

The performance of FLD hinges on collecting seed samples in a communication-efficient way while preserving data privacy. One simple solution is to collect a small fraction of local data samples. As shown in Fig.~5b with 10 devices, each device's uploading only 2\% of 500 local data samples yields the higher accuracy and faster training convergence of FLD compared to Vanilla FL and FD. More general methods of seed sample collection will be presented in Sec.~4.

\section{Surrogate Data Exchange}
This section focuses on two problems. One is to rectify non-IID datasets of devices (see \textbf{Q6}). Non-IID data degrades the gains from model parameter and model output exchange methods, since their benefits are rooted in exploiting the diversity and correlation of their training data. The other issue is to complete training in a single communication round (\textbf{Q7}). Minimizing communication rounds is useful not only for reducing communication bandwidth and energy, but also for mitigating the risk of privacy leakages incurred during information exchanges. To address these problems, we introduce two surrogate methods that are equivalent to exchanging raw data samples in a communication-efficient and privacy-preserving manner: by exchanging a summary of raw datasets (\textbf{S6}), or by collectively training a realistic synthetic sample generator~(\textbf{S7}).

\subsection{ Federated Data Summarization}

(\textbf{S6}) \emph{Data summarization} extracts useful information from local datasets at each device. This can be utilized for rectifying non-IID training data. Furthermore, by aggregating the data summary at a single location, ML models can be trained without further exchanging raw data samples. Compared to raw datasets, their data summaries usually have much smaller sizes and less information about each sample, thereby saving communication bandwidth while preserving privacy. There are three broad categories of data summarization techniques~\cite{Ko2019DAIS}.

\vspace{5pt}\noindent1) \textbf{Statistical Summary}: Basic forms of statistical summary are representative values, such as mean, median, sum, and variance. These values can be computed for every regular time interval, spatial region, or data class. More advanced forms of statistical summary include histograms and statistical models representing raw datasets. For histograms, as an example, bag-of-words ML is based on the occurrence of different words, which is represented as a histogram. For statistical model approaches, the Gaussian mixture model (GMM) is often utilized in acoustic recognition for determining which class each acoustic signal segment belongs to. 

\vspace{5pt}\noindent2) \textbf{Dimensionality Reduction}:
As exemplified in Sec. 3.1, raw data dimension can be reduced by quantizing states in reinforcement learning or grouping data samples into labels for classification tasks. More generally, it is useful to project high-dimensional raw data samples onto a low-dimensional space. In this direction, principle component analysis (PCA) is one popular way, representing each raw sample using a linear combination of orthogonal principle components. For non-linear dimensionality reduction, feature extraction is another well-known method. This step can be separately performed prior to ML, or integrated into a deep neural network whose low layers play the feature extracting role.

\vspace{5pt}\noindent3) \textbf{Data Subsampling}: Selecting a small amount of samples from the dataset also plays a role of data summarization. A straightforward approach is random sampling, which however does not provide any guarantee on how well the sampled subset represents the original dataset. Alternatively, \emph{coreset}~\cite{Lu:19} is a notable data subsampling technique, which provides theoretical error bounds when an ML model is trained using the coreset. For a stricter privacy requirement, there also exist methods of constructing coresets that preserve the differential privacy of data.
\vspace{5pt}

The above three categories are not necessarily independent of each other, but can be integrated into a single data summarization scheme. For instance, one can compute the coreset on features extracted from images, which is a combination of dimensionality reduction and data subsampling.

\begin{figure}\centering
\subfigure[MultFAug~\cite{Jeong:FML19}.]{\includegraphics[width=\columnwidth]{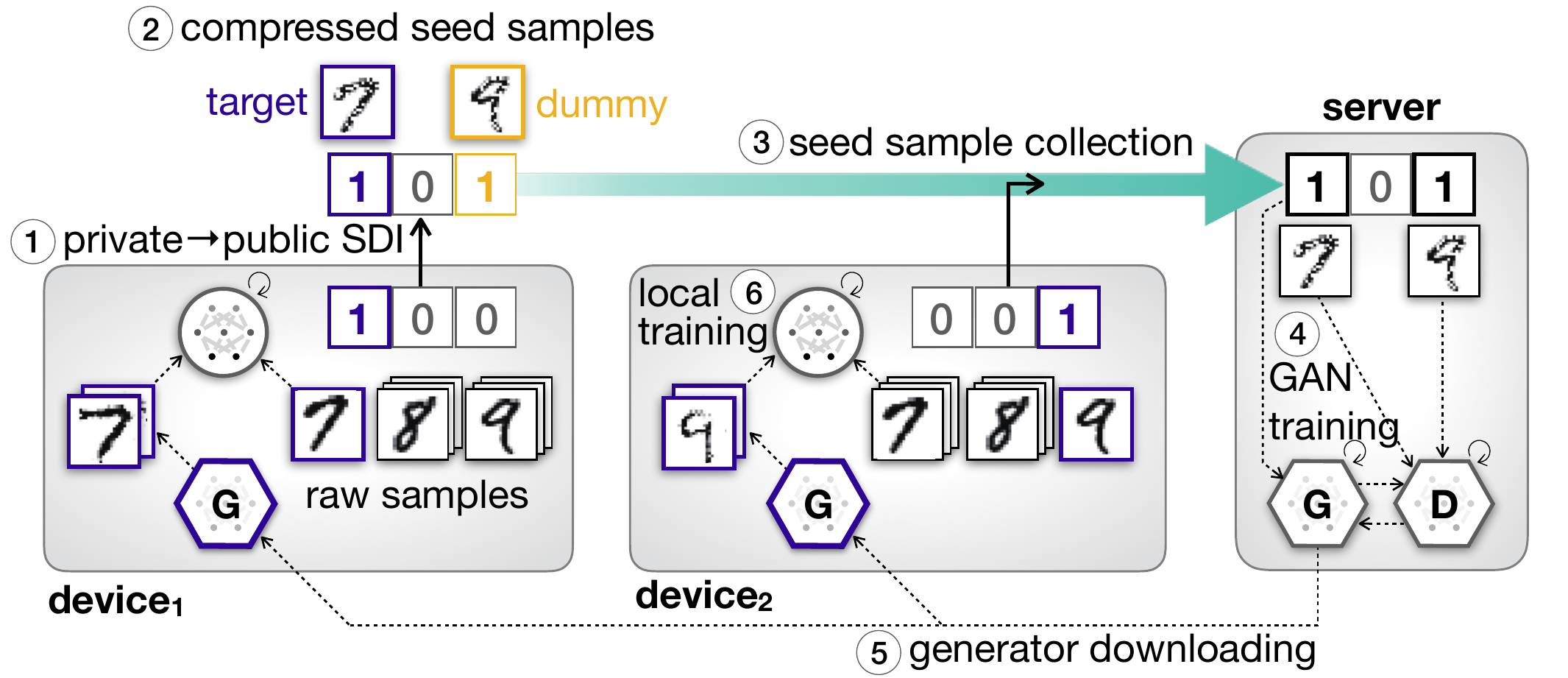}}
\subfigure[Accuracy of Vanilla FL with MultFAug~(left) and its label privacy guarantee (right, $1-\text{private SDI}/\text{aggregated public SDI}$) with $10$ or $25$ devices under non-IID MNIST datasets (target label: 4 samples; non-target label: 200 samples, per device).]{\includegraphics[width=.95\columnwidth]{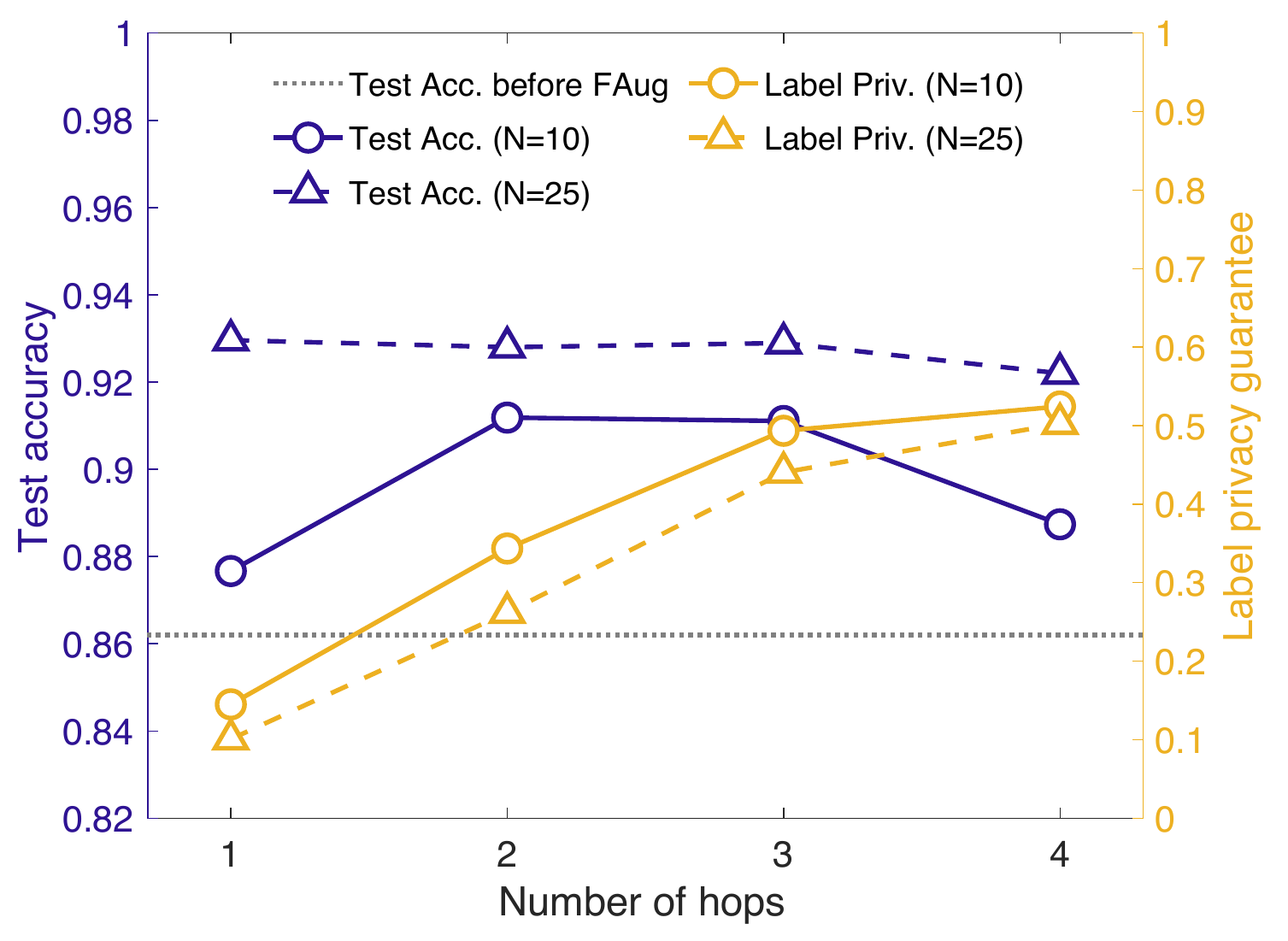}} \vskip -5pt
\caption{\emph{MultFAug} (Multi-hop Federated Data Augmentation with Sample Compression): Federated training of a surrogate data generator, while collecting compressed seed samples through multiple hops~\cite{Jeong:FML19}.}
\end{figure}

\subsection{Multi-Hop Federated Data Augmentation}

When devices need to exchange big data, it is more communication efficient to make them collectively train a data sample generator in a federated fashion. By sharing the sample generator, each device can locally generate synthetic data samples resembling the data samples stored in the other devices, without further communication. 

(\textbf{S7}) \emph{Federated data augmentation (FAug)} realizes this idea for an image classification task~\cite{Jeong:18}. Namely, in FAug (see Fig.~6a), devices collectively train a conditional generative adversarial network (GAN) stored at a server. The GAN consists of a discriminator neural network paired with another generator neural network to be downloaded by each device. These two neural networks are trained in a semi-supervised learning way, in which the discriminator evaluates the authenticity of the synthetic samples produced by the generator. To train the GAN, every device should upload: (i) a vector specifying its \emph{target labels} of lacking data samples; and (ii) \emph{seed samples} appended to each target label. 

These target label information and seed samples determine the communication payload sizes. Furthermore, they are often privacy sensitive, e.g., medical X-ray images (samples) are private, and the medical checkup items (target labels) may easily reveal diagnosis results. \emph{Multi-hop federated data augmentation with sample compression (MultFAug)}~\cite{Jeong:FML19} preserves such target label and sample privacy while maximizing communication efficiency as detailed next.

\vspace{5pt}\noindent1) \textbf{Multi-Hop Communication}: 
Target labels out of all labels are identified in a vector, referred to as sample distribution information (SDI). To preserve the target label privacy, the truthful SDI of each device, termed \emph{private SDI}, should be hidden from others. To this end, each device inserts dummy label indicators into the private SDI, yielding its \emph{public SDI}. For example, $\textsf{device}_1$ in Fig.~6a has the target label $\mathsf{7}$ of lacking samples. Then, $[1,0,0]$ is its private SDI for labels $\mathsf{7}$-$\mathsf{9}$. The public SDI is $[1,0,1]$ when a dummy label indicator is inserted into the label $\mathsf{9}$. 

Multi-hop communication makes the public SDI of devices be more truthful. In Fig.~6a, consider the next hop of $\textsf{device}_1$ is given as $\textsf{device}_2$ having its target label $\mathsf{9}$. The private SDI $[0,0,1]$ of $\textsf{device}_2$ can identically become the public SDI, by hiding its privacy in the preceding public SDI $[1,0,1]$ of $\textsf{device}_1$. Constructing truthful public SDI decreases dummy label indicators appended with redundant seed samples, thereby reducing the communication payload sizes. Furthermore, multi-hop communication can also provide path loss gains, particularly for faraway devices that cannot reach directly to the server.

\vspace{5pt}\noindent2) \textbf{Sample Compression}:
Every device under MultFAug compresses its seed samples by randomly removing a fraction of each sample. This sample compression reduces its privacy leakage. Indeed, more compressed samples become less similar to each other, when evaluating the similarity via the classical multidimensional scaling (MDS) algorithm. Moreover, the sample compression decreases communication payload sizes. This is viable by converting the compressed samples into compressed sparse row (CSR) formats. Each CSR representation only describes its original sample's non-zero elements with their row and column indices, thereby reducing the communication overhead.

The effectiveness of MultFAug on Vanilla FL under non-IID data is shown in Fig.~6b. For data privacy, the more hops, the better, allowing devices to hide their privacy more using less dummy labels, i.e., per-hop payload size reduction. For test accuracy, there exists an optimal number of hops. Indeed, more hops entail more transmission attempts, balanced by the per-hop payload reductions and path loss gains. This highlights the importance of network architecture and communication protocol designs, under the trade-off between privacy and communication-efficiency.

\section{Discussion and Conclusions}\label{sec:conclusion}

We conclude this article with a brief discussion about extending our proposed FML designs.

\vspace{5pt}\noindent 1) \textbf{Realistic {RAN} Characteristics}:
Wireless interference and channel fluctuations may perturb information exchanges, inserting undesired noise into ML operations. To address this, \emph{realistic channel dynamics} with \emph{antenna patterns} and their impact on \emph{interference} should be taken into account. Furthermore, different types of exchanges may have distinct latency and reliability requirements. Selecting their suitable \emph{RAN service links} and optimally \emph{multiplexing} them are important challenges. {Lastly, FML enables ML aided fog RAN, from which FML can also benefit. Therefore, it is necessary to \emph{jointly optimize FML and fog RAN operations}.}

\vspace{5pt}\noindent 2) \textbf{Heterogeneous Hardware Specifications}:
The available \emph{energy}, \emph{computing power}, and \emph{memory} size of every device are different from each other, and an order of magnitude smaller than the edge server's. Adaptive FL needs to account for these heterogeneous hardware specifications for optimizing local computing iterations and global communication rounds. Particularly under heterogeneous memory capacities, FD becomes more useful, which allows devices having different model sizes to collaborate with each other, as opposed to model parameter exchanging methods. Lastly, FLD and FAug transform the server's high computing power into the solutions to devices' limited transmission power and non-IID datasets, respectively. Depending on the \emph{computing power difference between the edge server and devices}, optimizing the server's role is an interesting problem.

\vspace{5pt}\noindent 3) \textbf{User Privacy and Fairness}:
User-generated data can preserve its privacy by distorting raw samples or adding dummy information in FML. Therefore, preserving more privacy without losing accuracy requires more federating devices with more communication rounds in a less energy-efficient way. In this respect, \emph{balancing privacy against communication and energy costs}, the aforementioned 1) and~2) should be co-designed. Finally, preceding FML frameworks serve devices in a best-effort way, and equally aggregate local updates. The resulting global information can be biased, since it more frequently reflects the local datasets of non-straggling devices having good channel conditions and high computing power. Possible solutions to this problem could \emph{fairly schedule} the devices straggled in consecutive communication rounds, and could minimize a \emph{fairness-aware loss function}, rather than the current empirical risk~minimization.

\bibliographystyle{ieeetr,IEEEabrv}  

\end{document}